\documentclass[12pt]{article}
\usepackage{subfigure}
\usepackage{amsmath,epsfig}

\usepackage{graphicx}
\usepackage{dcolumn}
\usepackage{bm}

\begin{document}

\title{\bf Identifying the underlying physics of the ridge via 3-particle $\Delta\eta-\Delta\eta$ correlations }

\author {Pawan Kumar Netrakanti (for the STAR Collaboration)\\ 
{\it Purdue University, USA.}\\
e-mail: pawan@purdue.edu}
\date{}

\maketitle

\begin{abstract}

We present the first results on 3-particle $\Delta\eta$-$\Delta\eta$ correlations
in minimum bias $d$+Au, peripheral and central Au+Au collisions at 
$\sqrt{{\it s}_{NN}}$ = 200 GeV measured by the STAR experiment. 
The analysis technique is described in
detail. The ridge particles, observed at large
$\Delta\eta$ in dihadron correlations in central Au+Au 
collisions, appear to be uniformly distributed over the measured 
$\Delta\eta$-$\Delta\eta$ region in 3-particle correlation. The results, together 
with theoretical models, should help further our understanding of the 
underlying physics of the ridge.

\vspace{0.4in}

\end{abstract}

\section{Introduction}

Dihadron correlations provide a powerful tool to study the properties of the
medium created in ultra-relativistic heavy-ion collisions. 
The observation of the near-side ridge in central Au+Au collisions~\cite{PRL,QM06}, where 
hadrons are correlated with a high transverse momentum ($p_{\perp}$) trigger particle in 
the azimuthal angle ($\Delta\phi\sim$0) but distributed approximately uniformly in 
pseudorapidity ($\Delta\eta$), has generated great interest. 
The properties of ridge particles, such as their $p_{\perp}$ spectral shape and 
particle compositions, are similar to those of inclusive particles, however 
the origin of the ridge is presently not understood.
Various theoretical models have been proposed, including 
longitudinal flow push~\cite{nestor}, QCD bremsstrahlung radiation boosted by 
transverse flow~\cite{voloshin,shuryak}, recombination between thermal and shower 
partons at intermediate $p_{\perp}$~\cite{hwa}, broadening of quenched jets in turbulent 
color fields~\cite{majumdar}, and elastic collision between hard and medium partons 
(momentum kick)~\cite{wong}. 
Production of correlated particles in all these models can be broadly divided into
two categories: (1) particles from jet fragmentation in vacuum which generate 
a jet-cone peak in dihadron correlation, and (2) particles
from gluon radiation affected by the medium and diffused broadly in $\Delta\eta$
which generate the ridge.
However, the qualitative features of dihadron correlations are all same
from these models. On the other hand, because the physics mechanisms of 
gluon diffusion in $\Delta\eta$ differs between models, the distribution of
two ridge particles in coincidence with the trigger particle can differ. 
Therefore, we analyze the 3-particle correlation in 
$\Delta\eta$-$\Delta\eta$ between two associated particles and a trigger particle
to potentially discriminate between the physics mechanisms proposed in these models. 
Jet fragmentation in vacuum would give a peak around 
($\Delta\eta_{1}$,$\Delta\eta_{2}$)$\sim$(0,0) in 3-particle $\Delta\eta$-$\Delta\eta$ 
correlations.
Particles from $\Delta\eta$ diffusion would produce structures that depend on the
physics mechanisms of diffusion and thus can be used to discriminate models. 
Combinations of one particle from jet fragmentation in vacuum and the other from $\Delta\eta$ 
gluon diffusion would generate horizontal or vertical strips in 3-particle 
$\Delta\eta$-$\Delta\eta$ correlations.

\section{Analysis technique and systematic uncertainties}

The data used in this analysis are from $d$+Au and Au+Au collisions at 
$\sqrt{{\it s}_{NN}}$ = 200 GeV and were taken by the STAR Time Projection Chamber 
(TPC)~\cite{tpc}. The Au+Au collisions were recorded with the minimum bais 
trigger and central trigger from zero degree calorimeters.    
The $z$-position of the constructed primary vertex (collision point) was restricted 
within $\pm$30 $\rm cm$ from the center of the TPC.
To ensure that these tracks come from the collision, the distance of closest approach to 
the primary vertex of less than 3.0 $\rm cm$ was used. The number of track points in the 
TPC was required to be greater than 15.
The trigger and associated particles are restricted to $\mid\eta\mid<$1 and their  
$p_{\perp}$ ranges are 3$<p_{\perp}^{trig}<$10 GeV/c and 1$<p_{\perp}^{assoc}<$3 GeV/c, respectively. 
The correlation yields are corrected 
for the centrality-, $p_{\perp}$-, $\phi$-dependent reconstruction efficiency for 
associated particles and the $\phi$-dependent efficiency for trigger particles,
and are normalized per corrected trigger particle.
\begin{figure*}[htp]
\includegraphics[height=13pc,width=38pc]{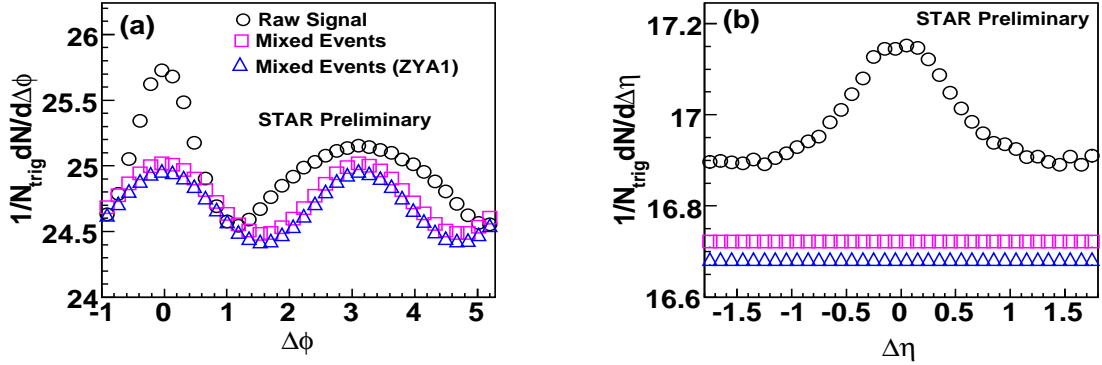}
\caption{
Two-particle $\Delta\phi$ and $\Delta\eta$ correlations in 
0-12$\%$ central Au+Au collisions are plotted in (a) and (b) respectively. 
The mixed event background (before and after $a$ scaling) are also shown.
Both the trigger and associated particles 
are restricted to $\mid\eta\mid<$1. The trigger and associated particles $p_{\perp}$ ranges are
 3$<p_{\perp}^{trig}<$10 GeV/c and 1$<p_{\perp}^{assoc}<$3 GeV/c, respectively.
The $\Delta\eta$ distributions 
are obtained for near-side associated particles within $\mid\Delta\phi\mid<$0.7. 
All $\Delta\eta$ correlations are corrected for the 2-particle $\Delta\eta$ acceptance.
}
\label{Fig1}
\end{figure*}

Figure~\ref{Fig1} (a) shows the 2-particle correlation signal in $\Delta\phi$ 
for 0-12$\%$ central Au+Au collisions, Y($\Delta\phi$). 
Also shown is the background constructed from the event-mixing technique, mixing a 
trigger particle from a triggered event with an associated particle from another 
event from the inclusive data sample. 
The inclusive event is required to have the same centrality, same magnetic 
field configuration and similar primary vertex $z$ position 
($\mid \Delta z \mid<$1 $\rm cm$) as for the triggered event.
The flow contribution is added by hand for the associated particle as it is 
not preserved in the mixed event background $B_{inc}(\Delta\eta,\Delta\phi)$.
The mixed event background is then scaled  by a constant $a$:
\begin{equation}
B(\Delta\phi) = \it{a}\int^{\rm 1}_{\rm -1}B_{\rm inc}(\Delta\eta,\Delta\phi)\left[1+{\rm 2}v_{\rm 2}^{\rm trig}v_{\rm 2}^{\rm assoc}\cos{({\rm 2}\Delta\phi)}+{\rm 2}v_{\rm 4}^{\rm trig}v_{\rm 4}^{\rm assoc}\cos{({\rm 4}\Delta\phi)}\right]d(\Delta\eta).
\label{eq0}
\end{equation}
This normalization was performed in the 0.8$<\Delta\phi<$1.2 range to match the correlation 
signal assuming zero yield at $\Delta\phi=$1 radian (ZYA1). 
The $v_{2}$ and $v_{4}$ are the anisotropic flow coefficients,  and are measured to
be independent of $\eta$~\cite{flow}.

Figure~\ref{Fig1} (b) shows the $\Delta\eta$ distribution within $\mid \Delta\phi \mid <$ 0.7 
on the near side. The background constructed 
from mixed event is scaled by the same $a$ factor as obtained by the 2-particle ZYA1 
in $\Delta\phi$:
\begin{equation}
B(\Delta\eta) = \it{a}\int^{\rm 0.7}_{\rm -0.7}B_{\rm inc}(\Delta\eta,\Delta\phi)\left[1+{\rm 2}v_{\rm 2}^{\rm trig}v_{\rm 2}^{\rm assoc}\cos{({\rm 2}\Delta\phi)}+{\rm 2}v_{\rm 4}^{\rm trig}v_{\rm 4}^{\rm assoc}\cos{({\rm 4}\Delta\phi)}\right]d(\Delta\phi).
\label{eq1}
\end{equation} 
The correlated 2-particle yield is given by:
\begin{equation}
\hat{Y}(\Delta\eta) = Y(\Delta\eta) - B(\Delta\eta).
\label{eqA}
\end{equation}
In Figure~\ref{Fig1} (b), the additional 2-particle $\Delta\eta$ acceptance, $A(\Delta\eta)$, is 
applied on both the signal, Y$(\Delta\eta)$/$A(\Delta\eta)$, and the background,
B$(\Delta\eta)$/$A(\Delta\eta)$.
The broad jet-like peak around $\Delta\eta\sim$0 is observed, and this peak 
sits atop of a relatively flat structure which represents the ridge. 

The 3-particle correlation raw signal, Y$(\Delta\eta_{1})\otimes$Y$(\Delta\eta_{2})$, is obtained 
from all triplets of one trigger particle and two associated particles from the same triggered event. 
The associated particles were constrained in azimuthal angle relative to 
the trigger particle within $\mid \Delta\phi \mid <0.7$. 
The signal is binned in $\Delta\eta_{1}$ and $\Delta\eta_{2}$, the pseudorapidity 
differences between the associated particles and the trigger. 
The raw 3-particle correlation signal can be formulated as:
\begin{eqnarray}
Y(\Delta\eta_1)\otimes Y(\Delta\eta_2) & = &\hat{Y}(\Delta\eta_1)\otimes \hat{Y}(\Delta\eta_2) + B(\Delta\eta_1)\otimes B(\Delta\eta_2) \notag\\ 
& & + \left[\hat{Y}(\Delta\eta_1)\otimes B(\Delta\eta_2) + \hat{Y}(\Delta\eta_2) \otimes B(\Delta\eta_1)\right]
\label{eq2}
\end{eqnarray}
where $\hat{Y}(\Delta\eta)$ and  B$(\Delta\eta)$ represent the correlated and 
background particles, respectively. 
The two sources of background in the raw 3-particle correlation are: 
(1) one of the two associated particles is correlated with the trigger particle 
besides flow correlation, and
(2) neither of the two associated particles is correlated with the trigger particle 
besides flow correlation.

The first background, referred to as Hard-Soft (HS), cannot be readily obtained from the folding 
of the background subtracted 2-particle correlation with the underlying background
because of the non-uniform 2-particle $\Delta\eta$ acceptance. The folding would result in
the product of two averages, the average 2-particle correlation, $\hat{Y}(\Delta\eta)$,
and the average background, B($\Delta\eta$).
Since $\hat{Y}(\Delta\eta)$ and B($\Delta\eta$) are correlated event-by-event because of the 
$\Delta\eta$ acceptance, the average of the product does not equal to the product of
the averages. Instead we construct the HS by mixing trigger-associated pair from the triggered event 
with a particle from a different and inclusive event.
Namely,
\begin{eqnarray}
HS & = & \hat{Y}(\Delta\eta_{1})\otimes B(\Delta\eta_{2}) + \hat{Y}(\Delta\eta_{2})\otimes B(\Delta\eta_{1})\notag\\
& = & a\Bigl[Y(\Delta\eta_{1})B_{\rm inc}(\Delta\eta_{2})\Bigr]F^{(2)} + a\Bigl[Y(\Delta\eta_{2})B_{\rm inc}(\Delta\eta_{1})\Bigr] F^{(1)} \notag\\ 
& & - 2a^2\Bigl[B_{\rm inc}(\Delta\eta_{1})B_{\rm inc}(\Delta\eta_{2})\Bigr]F.
\label{eq3}
\end{eqnarray}
Here the last term is constructed by mixing two different inclusive events to take care of the
uncorrelated part in the first two terms of the Eq.~\ref{eq3}.
The $F^{(1)}$ and $F^{(2)}$ are to take into account the flow correlation related to 
associated particle 1 and 2, respectively, and are given by:
\begin{eqnarray}
F^{(1)} & = & \langle 1+2v_2^{\rm trig}v_2^{(1)}\cos(2\Delta\phi_{1}) + 2v_2^{(1)}v_2^{(2)}\cos(2\Delta\phi_{1}-2\Delta\phi_{2}) + 2v_4^{\rm trig}v_4^{(1)}\cos(4\Delta\phi_{1})\notag\\
& & +2v_4^{(1)}v_4^{(2)}\cos(4\Delta\phi_{1}-4\Delta\phi_{2}) + 2v_2^{\rm trig}v_2^{(1)}v_4^{(2)}\cos(2\Delta\phi_{1}-4\Delta\phi_{2})\notag\\
& & +2v_2^{\rm trig}v_2^{(2)}v_4^{(1)}\cos(4\Delta\phi_{1}-2\Delta\phi_{2})+2v_2^{(1)}v_2^{(2)}v_4^{\rm trig}\cos(2\Delta\phi_{1}+2\Delta\phi_{2})\rangle 
\label{f32}
\end{eqnarray}
and an analogous equation for $F^{(2)}$ with $1\leftrightarrow2$.
The $F$ is to take into account the flow correlation among all the three particles 
in the event mixing, and is given by
\begin{eqnarray}
F & = & \langle 1+2v_2^{\rm trig}v_2^{(1)}\cos(2\Delta\phi_{1})+2v_2^{\rm trig}v_2^{(2)}\cos(2\Delta\phi_{2})+2v_2^{(1)}v_2^{(2)}\cos(2\Delta\phi_{1}-2\Delta\phi_{2})\notag\\
& & +2v_4^{\rm trig}v_4^{(1)}\cos(4\Delta\phi_{1})+2v_4^{\rm trig}v_4^{(2)}\cos(4\Delta\phi_{2})+2v_4^{(1)}v_4^{(2)}\cos(4\Delta\phi_{1}-4\Delta\phi_{2})\notag\\
& & +2v_2^{\rm trig}v_2^{(1)}v_4^{(2)}\cos(2\Delta\phi_{1}-4\Delta\phi_{2})+2v_2^{\rm trig}v_2^{(2)}v_4^{(1)}\cos(4\Delta\phi_{1}-2\Delta\phi_{2})\notag\\
& & +2v_2^{(1)}v_2^{(2)}v_4^{\rm trig}\cos(2\Delta\phi_{1}+2\Delta\phi_{2})\rangle.
\label{f33}
\end{eqnarray}
The averages in Eq.~\ref{f32},~\ref{f33} are taken within $\mid\Delta\phi_{1,2}\mid<$0.7.
The superscripts represent the $v_{2}$ and $v_{4}$ for trigger particle and
associated particles. We used a parameterization of $v_{4}$=1.15$v_{2}^{2}$.
Again, the flow contributions are constant in the measured $\eta$ range. 

The second background, referred to as Sof-Soft (SS), is constructed by mixing a trigger 
particle with an associated particle pair from an inclusive event which 
preserves all correlations between the two associated particles: 
\begin{equation}
SS=B(\Delta\eta_1)\otimes B(\Delta\eta_2)=a^2b\Bigl[B_{\rm inc}(\Delta\eta_{1})\otimes B_{\rm inc}(\Delta\eta_{2})\Bigr]F^{(t)}
\label{eq4}
\end{equation}
where $a$ is the same factor as obtained from 2-particle ZYA1 in $\Delta\phi$.
The flow contribution between trigger particle and the background 
particles is not preserved in the event mixing. This contribution, the so-called trigger flow, is
added by hand:
\begin{eqnarray}
F^{(t)} & = & \langle 1+2v_2^{\rm trig}v_2^{(1)}\cos(2\Delta\phi_{2})+2v_2^{\rm trig}v_2^{(2)}\cos(2\Delta\phi_2)+2v_4^{\rm trig}v_4^{(1)}\cos(4\Delta\phi_{1})\notag\\
& &  +2v_4^{\rm trig}v_2^{(2)}\cos(4\Delta\phi_{2})+2v_2^{\rm trig}v_2^{(1)}v_4^{(2)}\cos(2\Delta\phi_{1}-4\Delta\phi_{2})\notag\\
& & +2v_2^{\rm trig}v_2^{(2)}v_4^{(i)}\cos(4\Delta\phi_{1}-2\Delta\phi_{2})+2v_2^{(1)}v_2^{(2)}v_4^{\rm trig}\cos(2\Delta\phi_{1}+2\Delta\phi_{2})\rangle
\label{eqB}
\end{eqnarray}
where the average is taken within $\mid\Delta\phi_{1,2}\mid<$0.7.

The factor $a^{2}b$ in Eq.~\ref{eq4} scales the number of associated pairs from the inclusive event to that
in the background underlying the triggered event: 
\begin{equation}
b = \frac{\Bigl[\langle N_{\rm assoc}(N_{\rm assoc}-1) \rangle/\langle N_{\rm assoc} \rangle^2\Bigr]_{\rm bkgd}}{\Bigl[\langle N_{\rm assoc}(N_{\rm assoc}-1) \rangle/\langle N_{\rm assoc} \rangle^2\Bigr]_{\rm inc}}
\label{eqC}
\end{equation} 
where $N_{\rm assoc}$ denotes the associated particle multiplicity.
If the associated particle multiplicity in the inclusive
event and in the background underlying the triggered event are both Poissonian or deviate 
from Poissonian equally, then $b$=1.
In our analysis, we obtain $b$ in the following way. We scale the 2-particle $\Delta\eta$ 
distribution such that the ridge contribution in 1.0$<\mid \Delta\eta \mid<$1.8 is zero, 
and this gives a new $a$. We repeat our analysis with this new $a$, and obtain
$b$ by requiring the average 3-particle $\Delta\eta$-$\Delta\eta$ signal 
in 1.0$<\mid (\Delta\eta_{1},\Delta\eta_{2}) \mid<$1.8 to be zero.
We use the thus obtained $b$ in our analysis with the default $a$ to obtain the final
3-particle correlation.
The assumption in this is:
\begin{equation}
\Bigl[\langle N_{\rm assoc}(N_{\rm assoc}-1)\rangle/\langle N_{\rm assoc} \rangle^2\Bigr]_{\rm bkgd} = \Bigl[\langle N_{\rm assoc}(N_{\rm assoc}-1)\rangle/\langle N_{\rm assoc} \rangle^2\Bigr]_{\rm bkgd+ridge}.
\end{equation}
The 3-particle $\Delta\eta$-$\Delta\eta$ correlation, $\hat{Y}(\Delta\eta_{1})\otimes\hat{Y}(\Delta\eta_{2})$, is obtained by subtracting the HS ans SS backgrounds from the raw signal. 
The obtained correlation is corrected for 3-particle $\Delta\eta$ acceptance. 
The acceptance is obtained from event-mixing of a 
trigger particle with associated particles from two different inclusive events, as was done for the
last term in Eq.~\ref{eq3}, namely
\begin{equation}
A(\Delta\eta_1,\Delta\eta_2) = \frac{B_{\rm inc}(\Delta\eta_1) B_{\rm inc}(\Delta\eta_1)}{B_{\rm inc}(0) B_{\rm inc}(0)}.
\label{eq10}
\end{equation}

The main sources of systematic uncertainty in our 3-particle correlation results 
are from the normalization factors $a$, $b$ and the flow measurements.
The $v_{2}$ used in our analysis is the average $v_{2}$ from the modified reaction
plane method and the 4-particle cummulant method~\cite{PRL}. We assign a $\pm$10$\%$ systematic
uncertainty on $v_{2}$.
The systematic uncertainty on $a$ is estimated by using the normalization
range of 0.9$<\Delta\phi<$1.1 and 0.7$<\Delta\phi<$1.3. 
The systematic uncertainty on $b$ is estimated by using the normalization 
range in $\Delta\eta$ of $\mathrm{-}$1.8$<\mid\Delta\eta\mid<\mathrm{-}$1.2 
and $\mathrm{-}$1.2$<\mid\Delta\eta\mid<\mathrm{-}$0.6. 

\section{Results and Discussion}
\begin{figure}[hp]
\includegraphics[height=13pc,width=38pc]{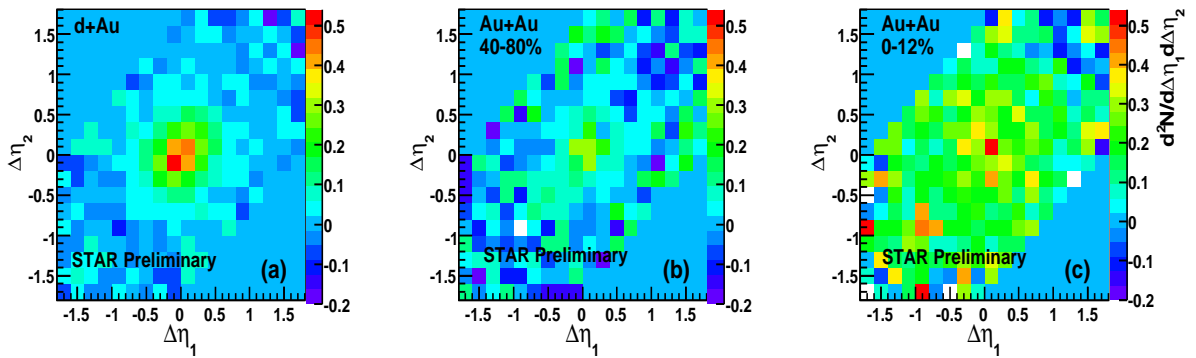}
\caption{Background subtracted 3-particle $\Delta\eta$-$\Delta\eta$ correlation in 
(a)$d$+Au, (b)40-80$\%$ Au+Au and (c) central 0-12$\%$ for Au+Au collisions. 
The trigger and associated particles $p_{\perp}$ ranges are 3$<p_{\perp}^{trig}<$10 GeV/c and 1$<p_{\perp}^{assoc}<$3 GeV/c, respectively.
The $\Delta\eta$-$\Delta\eta$ correlations are obtained for near-side associated 
particles within $\mid\Delta\phi\mid<$0.7. All correlations are corrected for the 
3-particle $\Delta\eta$ acceptance.
}
\label{Fig2}
\end{figure}

Figure~\ref{Fig2} (a), (b) and (c) show the background subtracted 3-particle 
$\Delta\eta$-$\Delta\eta$ correlation for $d$+Au, 40-80$\%$ Au+Au and 0-12$\%$ Au+Au 
at $\sqrt{{\it s}_{NN}}$ = 200 GeV, respectively.
The prominent jet structure is observed in $d$+Au and 40-80$\%$ Au+Au collisions
around ($\Delta\eta_{1}$,$\Delta\eta_{2}$)$\sim$(0,0). 
The peak is also observed in 0-12$\%$ central Au+Au collisions, but the peak is
atop of an overall pedestal. This pedestal is caused by the ridge particles, and does not
seem to have other structures in ($\Delta\eta_{1},\Delta\eta_{2}$).
The ridge particles seem to be distributed approximately uniformly over 
the measured $\Delta\eta$-$\Delta\eta$ region.

\begin{figure}[htp]
\centering
\includegraphics[height=18pc,width=26pc]{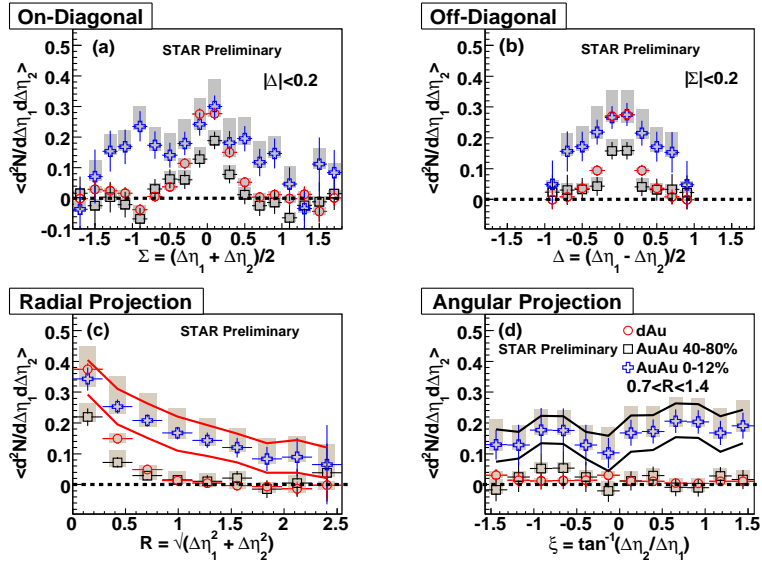}
\caption{Projections of the background subtracted 3-particle $\Delta\eta$-$\Delta\eta$ correlations 
in $d$+Au, 40-80$\%$ Au+Au and 0-12$\%$ central Au+Au collisions. Panels  (a)-(d) show 
the on-diagonal, off-diagonal, radial and angular projections, respectively. 
The shaded box represent the systematic uncertainty due to background normalization 
and the solid lines represents the systematic uncertainty due to flow subtraction.}
\label{Fig3}
\end{figure}

To study the $\Delta\eta$-$\Delta\eta$ correlation in more detail, 
Figure~\ref{Fig3} (a) and (b) show the projections of the 3-particle $\Delta\eta$-$\Delta\eta$ 
correlation along the on-diagonal $\Sigma=(\Delta\eta_{1} + \Delta\eta_{2})/2$ and
off-diagonal $\Delta=(\Delta\eta_{1} \mathrm{-} \Delta\eta_{2})/2$. 
These projections are performed within $\mid\Delta\mid<$0.2 and $\mid\Sigma\mid<$0.2, respectively.
Figure~\ref{Fig3} (c) shows the radial 
$R=\sqrt{(\Delta\eta_{1})^{2} + (\Delta\eta_{2})^{2}}$ projection. All projections
are normalized by the projected area, therefore they are the average correlation
signal per $\rm radian^{2}$.
The average signals peak at $\Sigma\sim$0 or $\Delta\sim$0, in $d$+Au and 40-80$\%$ Au+Au
collisions and rapidly fall off to zero at large $\Sigma$ or $\Delta$. For central 0-12$\%$ Au+Au 
collisions the signal also peaks at $\Sigma\sim$0 and $\Delta\sim$0 and is broadly distributed.
The radial projection is peaked at $R\sim$0 and drops gradually with R and 
perhaps flattens out in central 0-12$\%$ Au+Au collisions.
Figure~\ref{Fig3} (d) shows the angular projection 
$\xi$=$tan^{-1}$($\Delta\eta_{2}$/$\Delta\eta_{1})$ in 0.7$<R<$1.4. The average signal over $\xi$ 
shows no evidence for horizontal or vertical strips in the $\Delta\eta$-$\Delta\eta$ correlation.

Our results suggest that the ridge particles in the central 0-12$\%$ Au+Au collisions 
are uncorrelated in $\Delta\eta$ between themselves. The ridge appears to be uniform 
event-by-event.
We also observe the small $\Delta\eta$ peak, suggesting contributions from jet fragmentation
in vacuum. However, the two contributions, one from jet fragmentation in vacuum, 
the other from the ridge, do not seem to co-exist in the same event because we do not observe
the horizontal and vertical strips in $\Delta\eta$-$\Delta\eta$ correlation. Since no
plausible physics excludes the co-existence of these two effects, our results indicate that
the probability of such a co-existence is small.

\section{Summary and outlook}
We have presented the first results on 3-particle $\Delta\eta$-$\Delta\eta$ correlation 
for $d$+Au, 40-80$\%$ Au+Au and 0-12$\%$ Au+Au collisions at $\sqrt{{\it s}_{NN}}$ = 200 GeV. 
A correlation peak at ($\Delta\eta_{1}$, $\Delta\eta_{2}$)$\sim$(0,0) characteristic of 
jet fragmentation in vacuum, is observed in all systems.
This peak sits atop of a pedestal in central 0-12$\%$ Au+Au collisions. This pedestal, 
composed of particle pairs from the ridge, is approximately uniform or broadly falling 
within the measured $\Delta\eta$ 
acceptance. No other significant structures, except that from jet fragmentation in vacuum, 
were observed in the projection. The ridge particles are uncorrelated
among themselves in $\Delta\eta$. The ridge is uniform event-by-event.

To understand the physics mechanism(s) generating the ridge, quantitative model
calculations are clearly needed. 
These results in comparison to theoretical models and high statistic 
data sets will help furhter our understanding of the underlying physics
of the ridge.

\end{document}